# Detecting the Classical Harmonic Vibrations of Micro Amplitudes and Low Frequencies with an Atomic Mach-Zehnder Interferometer

Yong-Yi Huang†

*MOE Key Laboratory for Nonequilibrium Synthesis and Modulation of Condensed Matter,*
*Department of Optic Information Science and Technology,*
*Xi'an Jiaotong University, Xi'an 710049, China*

**Abstract**

We study the effects of atomic beams classical harmonic vibrations of micro amplitudes and low frequencies perpendicular to the wave vectors of atomic branches on the mean numbers of atoms arriving at the detectors in an atomic Mach-Zehnder interferometer, where the two atomic beams are in the same wave surface and have the same phase. We propose a vibrant factor $F$ to quantitatively describe the effects of atomic beams vibrations. It shows that: (i) the vibrant factor $F$ depends on the relative vibrant displacement and the initial phase rather than the absolute amplitude, (ii) the factor $F$ increases with the increase of the initial phase, and (iii) the frequencies can be derived from equal time interval measurements of the mean numbers of atoms arriving at the detectors. These results indicate that it is possible to detect the classical harmonic vibrations of micro amplitudes and low frequencies by measuring the variations in the mean numbers of atoms arriving at the detectors.

## I. Introduction

Atoms in an atom interferometer are deliberately given the option of traversing an apparatus through two or more alternate paths, and then an interference pattern is observed. Atom interferometers do not need a complex source and are far more sensitive than neutron interferometers because of their considerably large signal. The atom translational motions are coherently manipulated, where *coherently* means *based on the phase of de Broglie wave*. Atom interferometers already works with magneto-optical traps or Bose-Einstein condensates, so atom interferometers are useful tools for studying quantum mechanical phenomena, probing atomic and material properties and measuring inertial displacements[1,2,3,4]. Atom interferometer is even designed for the gravitational wave detection[5,6]. A Mach-Zehnder interferometer is a device to determine the relative phase shift between two collimated beams from a coherent light source[7,8], for instance one of the two beams is caused by a small sample or the change in length. An atomic Mach-Zehnder interferometer[9,10,11] is the simplest atom interferometer, the interferometers with mechanical gratings have been used to measure phase due to rotations[12], to study decoherence due to scattering photons and background gas[13,14,15], to perform the separated beam experiments with molecules(Na2)[11] and to measure the polarizability of He and He2[16]. The atomic Mach-Zehnder interferometer with light gratings has been used to measure the polarizability of Li atoms[17,18,19]. In an atomic Mach-Zehnder interferometer, an auxiliary interaction grating is inserted and removed from each path to measure the phase shift due to van der Waals interaction[20].

As we know, a gravitational wave is an extremely weak wave. A plane gravitational wave with two polarization states travelling in the z direction will deform the particles around a circle in the xy plane[21], a gravitational wave can be confirmed by measuring the oscillatory motions hit by a plane gravitational wave[22] (unfortunately, it is not accepted). In this article we study the detections of the classical harmonic vibrations of micro amplitudes and low frequencies. The classical harmonic vibrations are simpler but more fundamental than the oscillatory motions induced by a plane gravitational wave with two polarization states. We investigate the effects of atomic beams classical transverse harmonic vibrations of micro amplitudes and low frequencies on the mean numbers of atoms arriving at the detectors in an atomic Mach-Zehnder interferometer. We expect that the classical harmonic vibrations of micro amplitudes and low frequencies can be detected by measuring the variations of the mean numbers of atoms arriving at the detectors. Our methods probably provide a new detection principle for a gravitational wave. The paper is organized as follows. In section II we present how the atomic beams classical harmonic vibrations of micro amplitudes and low frequencies affect the mean numbers of atoms arriving at the detectors and derive a vibrant factor $F$ to quantitatively describe the situation. In section III we discuss the main properties of the vibrant factor $F$ and illustrate how to detect the classical harmonic vibrations by measuring the variations of the mean numbers of atoms arriving at the detectors. In section IV we give a summary.

**II. The Corrected Mean Numbers of Atoms Arriving at the Detectors due to the Atomic Beams Classical Transverse Harmonic Vibrations**

We study a thought experiment whereby a stream of atoms are sent through a Mach-Zehnder interferometer during a measurement time $\Delta t$, the two branches are in the paper plane, as shown in Fig. 1. The atoms are split at the beam-splitter 1, follow the paths $\alpha$ or $\beta$, are reflected off the mirrors, and are recombined at the beam-splitter 2. The atomic branches during the measurement time $\Delta t$ are induced by some plane harmonic wave, which makes the atomic branches vibrate perpendicular to the atomic wave vectors. To realize the expected design, we make a plane harmonic longitudinal wave perpendicular to the paper plane hit the two atomic branches. As for a polarized plane harmonic transverse wave, we make the transverse wave perpendicular to the paper plane be transformed into the shapes of the four branches, and now the vibrant directions induced by the transverse wave are in the paper plane and perpendicular to the wave vectors. The reasons why the vibrant directions of the branches are perpendicular to the wave vectors of the branches are that the collective classical harmonic vibrations of the branches induced by the plane harmonic wave fully decouple with the translational motions of the branches and that we can extract the frequency knowledge of the collective vibrations from the translational motions. How do we succeed? Please see the following text.

We suppose that the measurement time $\Delta t$ is much smaller than the harmonic vibration period. The recombined atoms are detected at the upper detector A or the lower detector B. The atoms are reflected from a beam-splitter surface, they have a phase shift. Without loss of generality, we take the phase shift to be $\pi/2$, one should know that the actual phase shift depends on the structure of the beam-splitter. After passing through a beam-splitter, the atom undergoes a phase shift $\varphi_i$ (i=1,2 for the first and the second beam-splitter respectively), the cumulative effect leads to a wave function $\psi_a, \psi_b$ corresponding to detector A and detector B[23],

$$\psi_a = \frac{\psi}{2} e^{i\theta_a}[1 - e^{-ik(l_\alpha - l_\beta)}] \qquad (1)$$

and

$$\psi_b = \frac{\psi}{2} e^{i\theta_b}[1 + e^{-ik(l_\alpha - l_\beta)}], \qquad (2)$$

where $\theta_a = kl_\alpha - \pi/2$ and $\theta_b = kl_\alpha$, and we let $\varphi_1 = \varphi_2 = \pi$. In Equ.(1) and Equ.(2), $k$ is the atomic wave vector, and $l_\alpha, l_\beta$ are the path lengths through the $\alpha$ and $\beta$ branches. We suppose the flux of the atomic stream is $j$ (the number of atoms per second passing through a plane), so the number of atoms in the stream during a measurement time $\Delta t$ is $N = j \cdot \Delta t$. An appropriate state vector for $i$th atom after the recombination by beam-splitter 2 is a superposition state and is given by

$$|\varphi>_i = \frac{e^{i\theta_a}}{2}(1 - e^{-i\varphi_{\alpha\beta}})|1_a, 0_b>_i + \frac{e^{i\theta_b}}{2}(1 + e^{-i\varphi_{\alpha\beta}})|0_a, 1_b>_i, \qquad (3)$$

where we set $\varphi_{\alpha\beta} \equiv k(l_\alpha - l_\beta)$, and $|1_a, 0_b>, |0_a, 1_b>$ denote an atom incident on detector A or detector B. The state vector of the atomic stream with the number $N$ of atoms is constructed by a direct product of the individual atomic states, i.e.

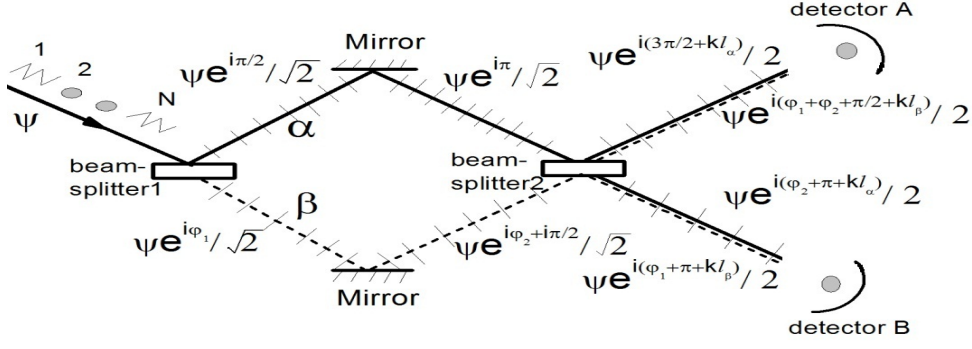

Fig. 1 Schematic of an atomic Mach-Zehnder interferometer, where phase changes by beam-splitters and mirrors illustrate accumulated phase shifts in $\alpha$ or $\beta$ branches, the short solid lines of the atomic branches denote the collective vibrations induced by a polarized plane harmonic transverse wave. The two atomic branches are in the same wave surface and have the same phase. The collective vibrant directions are perpendicular to the wave vectors of the two branches.

$$|\Phi>_N \equiv \prod_{i=1}^{N} |\varphi>_i .$$

Now we know that the atomic streams have a collective harmonic vibration with the same phase induced by some plane harmonic wave and that the vibrant directions are perpendicular to the wave vectors of the beams, so the collective vibrations do not couple with the translational motions. The Hamiltonian of atomic streams with the number $N$ of atoms is given by

$$H = \frac{P_\perp^2}{2M} + \frac{1}{2}M\Omega^2 x^2 + \sum_{i=1}^{N} \frac{p_i^2}{2m}, \qquad (4)$$

where $M = Nm$, $m$ is the mass of one atom, and $\Omega$ is the angle frequency of the plane harmonic wave. The state vector of the atomic streams is written as

$$|\Psi>_N = |n> \otimes |\Phi>_N = |n> \otimes \prod_{i=1}^{N} |\varphi>_i \quad (5)$$

In Equ.(5) $|n>$ is the eigenstate vector of the atomic stream vibrations, the corresponding eigen energy is $E_n = (n+1/2)\hbar\Omega$. In this article we suppose that $\Omega$ is so low that the behaviors of the harmonic vibrations can be regarded as those of a classical harmonic oscillator.

Let $c_{\sigma,i}^\dagger$ and $c_{\sigma,i}$, where $\sigma = A, B$, be the creation and annihilation operators for the number states $|n_A, n_B>_i$, and we have the number operators $n_{\sigma,i} = c_{\sigma,i}^\dagger c_{\sigma,i}$, where the eigenvalues $n_A$ and $n_B$ are 0 or 1. The operator $c$ obeys the commutation relations $c_{\sigma i} c_{\sigma j}^\dagger \pm c_{\sigma j}^\dagger c_{\sigma i} = \delta_{ij}$, where the plus or minus sign indicates Bose or Fermi statistics. However, the statistics are neglected in this article, because we suppose that the density of atomic stream is not large and there is only one atom at a time within a single coherence length. The number operator $N_\sigma$ of the atomic stream is given by

$$N_\sigma = \sum_{i=1}^{N} n_{\sigma i} \qquad (\sigma = A, B). \qquad (6)$$

The expectation values $<N_\sigma>_N$ of these number operators Equ.(6) are written as

$$_N<\Psi|N_A|\Psi>_N = \int_{\Delta t} <n|x><x|n>dx \cdot \sum_{i=1}^{N} \left|\frac{1-e^{-i\varphi_{\alpha\beta}}}{2}\right|^2 {}_i<1_A 0_B|n_{Ai}|1_A 0_B>_i$$

$$_N<\Psi|N_B|\Psi>_N = \int_{\Delta t} <n|x><x|n>dx \cdot \sum_{i=1}^{N} \left|\frac{1+e^{-i\varphi_{\alpha\beta}}}{2}\right|^2 {}_i<0_A 1_B|n_{Bi}|0_A 1_B>_i .$$

One easily works out the mean numbers of atoms in detector A and detector B during the measurement time $\Delta t$ [23],

$$_N<\Psi|N_A|\Psi>_N = N\sin^2\frac{\varphi_{\alpha\beta}}{2}\cdot\int_{\Delta t}<n|x><x|n>dx \qquad (7)$$

$$_N<\Psi|N_B|\Psi>_N = N\cos^2\frac{\varphi_{\alpha\beta}}{2}\cdot\int_{\Delta t}<n|x><x|n>dx, \qquad (8)$$

where $N = j\cdot\Delta t$ has above been declared. Because the measurement time $\Delta t$ is much smaller than the vibration period $T = 2\pi/\Omega$, the mean numbers of atoms in detector A or B have a correction $F \equiv \int_{\Delta t}<n|x><x|n>dx$.

We define the corrected coefficient $F$ of the mean numbers of atoms arriving at the detectors as the vibrant factor $F$, because it comes from the vibrations induced by the plane harmonic wave. The vibrant factor $F$ quantitatively describes the effects of the collective vibrations on the mean numbers of atoms in the detectors and includes the frequency knowledge of the collective vibrations. From the vibrant factor $F$, it becomes probable to confirm the existence of the collective vibrations, especially the vibrations with very micro amplitudes, by measuring the variations of the mean numbers of atoms arriving at the detectors. Now we derive the vibrant factor $F$. Because of the condition of low frequencies, $<n|x><x|n>$ can be regarded as a classical harmonic oscillator probability density. Given $\alpha = \sqrt{M\Omega/\hbar}$ and $\xi = \alpha x$, we obtain the classical vibrant equation of the atomic branches $\xi = \sqrt{2n+1}\sin(\Omega t + \delta)$, where $n$ is the vibrant quantum number, $\delta$ is the initial phases of the two atomic beams, and $\sqrt{2n+1}$ denotes the absolute amplitude. The classical harmonic oscillator probability density is $w(\xi) = <n|\xi><\xi|n> = \frac{1}{\pi\sqrt{(2n+1)-\xi^2}}$, and $w(\xi)$ increases with the increase of the displacement $\xi$ in $\xi\in[0,\sqrt{2n+1}]$ during one-half period $T/2$. We do not consider $\xi\in[-\sqrt{2n+1},0]$ region because of the probability density's symmetry between $[0,\sqrt{2n+1}]$ and $[-\sqrt{2n+1},0]$, seen from the probability density $w(\xi) = \frac{1}{\pi\sqrt{(2n+1)-\xi^2}}$.

We work out $F = \int_{\Delta t}<n|x><x|n>dx = \int_{\Delta t}<n|\xi><\xi|n>d\xi$ during a very short measurement time $\Delta t$ i.e.

$$F(\delta) = \int_{\xi_0}^{\xi_0+\zeta} w(\xi)d\xi = \frac{1}{\pi}[\arcsin(\frac{\zeta}{\sqrt{2n+1}}+\sin\delta)-\delta], \qquad (9)$$

where $\zeta$ denotes the absolute displacement during the measurement time $\Delta t$, and $\delta$ is the initial phase i.e. $\xi_0 = \sqrt{2n+1}\sin\delta$ with $\xi_0$ denoting the initial displacement at the initial time $t_0 = 0$. Equ. (9) is a fundamental formula in this article, which was firstly derived in [24]. It is the condition of the measurement time $\Delta t \ll T$ that makes the integral upper limit and lower limit in Equ.(9) be $[\xi_0, \xi_0+\zeta]$ rather than $(-\infty, +\infty)$ and lets the relative displacement be very small i.e. $0 < \frac{\zeta}{\sqrt{2n+1}} \ll 1$.

**III. the Vibrant Factor *F* and the Detection of the Classical Harmonic Vibrations of Micro Amplitudes and Low Frequencies with an Atomic Mach-Zehnder Interferometer**

The factor *F* versus the initial phase $\delta$ is shown in Fig. 2 with the relative displacement $\zeta/\sqrt{2n+1} = 0.05$. Here $\zeta/\sqrt{2n+1} = 0.05$ is not a particular choice but just an example, that is to say, $\zeta/\sqrt{2n+1}$ can be 0.02, 0.08, 0.1, 0.3 etc, the maximum of the relative displacement is unity. Actually we study the vibrant factor *F* during one period and require the measurement time $\Delta t$ be less than one period *T*, the shorter the measurement time $\Delta t$ is, the less the relative displacement $\zeta/\sqrt{2n+1}$ is. Seen from Equ.(9) and Fig.2 we observe: (1) The vibrant factor *F* depends on the relative vibrant displacement $\frac{\zeta}{\sqrt{2n+1}}$ and the initial phase $\delta$, rather than the absolute vibrant amplitude $\sqrt{2n+1}$. This result makes it possible to detect the vibrations of extremely micro amplitudes by measuring the variations of the mean numbers of atoms. (2) The vibrant factor *F* increases with the increase of the initial phase $\delta$. The reason is that the different initial phases $\delta$ corresponds to the different displacements $\xi$ and the probability density $w(\xi)$ increases with the increase of displacement $\xi$. We know that the point (2) is correct from Equ. (9).

During the measurement time $\Delta t \ll T$ the small relative displacement $\frac{\zeta}{\sqrt{2n+1}}$ is much smaller than unity, and the integral upper limit and lower limit in Equ.(9) is $[\xi_0, \xi_0+\zeta]$ rather than $(-\infty, +\infty)$, so the vibrant factor *F* is less than unity. What's more, the vibrant factor *F* greatly reduces the mean numbers of atoms in the two detectors by two orders of magnitude, seen from Equ. (7) and Equ. (8). Given $\dfrac{\zeta}{\sqrt{2n+1}}=0.05$, we maximally have $\sin\delta=0.95$, i.e. $\delta \simeq 1.25$, due to the fact $\dfrac{\zeta}{\sqrt{2n+1}}+\sin\delta=1$. Substituting $\delta \simeq 1.25$ into $F(\delta)$, we obtain that $F(\delta)$ maximum is about 0.1. If $\dfrac{\zeta}{\sqrt{2n+1}}=1$ is satisfied in Equ.(9), we have $\sin\delta=0$. The maximal value of $F(\delta)$ is 0.5, which corresponds to one-half period. In a whole period we have $F(\delta)=1$, however, it is a trivial result because we can not extract any useful information.

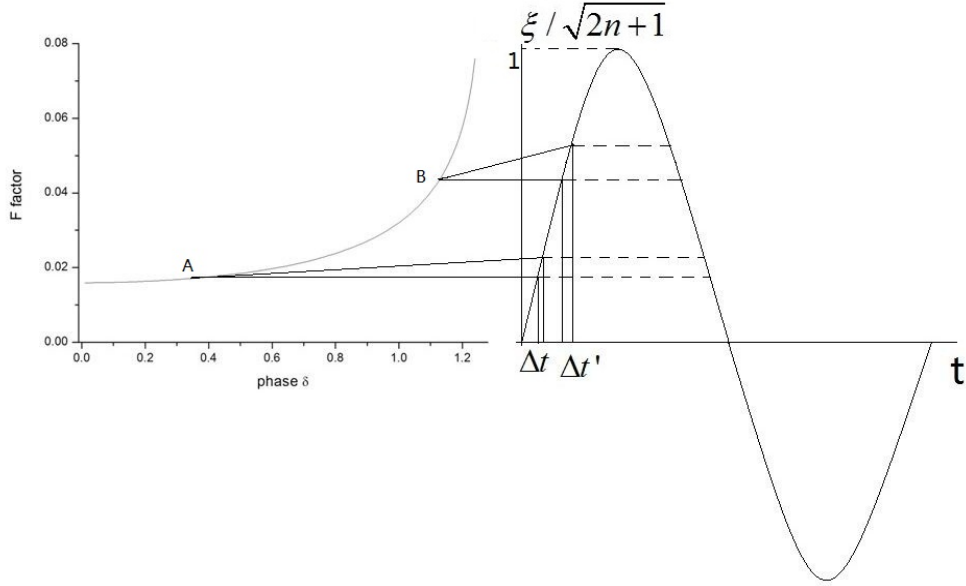

Fig. 2 The vibrant factor $F$ versus the initial phase $\delta$ with the relative displacement $\zeta/\sqrt{2n+1}=0.05$ lies in the left panel, and the right panel corresponding to the vibrant factor $F$ is the harmonic amplitude with time. A dot in the left panel corresponds to a measurement time $\Delta t$ with an initial phase $\delta$, and it takes different measurement time $\Delta t$ at different phase $\delta$ to keep the relative displacement $\zeta/\sqrt{2n+1}$ to be constant. For instance, dot A and dot B in the left panel have different initial phases and correspond to different displacements in the right panel. If the relative displacement $\zeta/\sqrt{2n+1}$ is kept to be constant (the measurement time in dot A is less than the time in dot B i.e. $\Delta t < \Delta t'$), the vibrant factor $F(A)$ is less than $F(B)$ because of the probability in dot A less than the probability in dot B seen from Equ.(9).

Given $\dfrac{\zeta}{\sqrt{2n+1}}=0.05$ in the detection process of the numbers of atoms, the mean numbers of atoms arriving at the detector A or detector B are given by

$$_N<\Psi|N_A|\Psi>_N = N\sin^2\frac{\varphi_{\alpha\beta}}{2}\cdot F(\delta)$$

$$_N<\Psi|N_B|\Psi>_N = N\cos^2\frac{\varphi_{\alpha\beta}}{2}\cdot F(\delta),$$

where $F(\delta) = \frac{1}{\pi}[\arcsin(0.05+\sin\delta)-\delta]$ and $N = j\cdot\Delta t$. Seen from Fig. 2, the larger the initial phase $\delta$ becomes, the longer measurement time $\Delta t$ it takes to keep the same relative displacement $\zeta/\sqrt{2n+1} = 0.05$ to be constant. Maybe it is very difficult and not practical to use this method to measure the mean numbers of atoms arriving at the detectors A or B, however, we realize that the vibrant factor $F$ increases with the increase of the initial phase and has nothing to do with the absolute vibrant amplitude, as shown in Fig.2. This fact implies that the measurement of the mean number variations of atoms arriving at detectors is one of many options to verify weak vibrations induced by a plane harmonic wave, no matter how small the vibrant amplitudes are. Theoretically once the left curve $F$ versus the phase is obtained, the micro classical harmonic vibrations of atomic beams are verified.

The other practical measurement of the mean numbers of atoms arriving at detectors is equal time interval measurements, i.e. we keep each measurement time $\Delta t$ to be constant. As shown in Fig.2, a relative vibrant displacement $\frac{\zeta}{\sqrt{2n+1}}$ contains two to-and-fro processes in one-half period, so we need to divide the vibrant factor $F(\delta)$ in Equ.(9) by 2. Let the measurement time $\Delta t$ be constant, we surprisedly obtain $\frac{F(\delta)}{2} = \frac{\Delta t}{T}$ independent of the initial phase $\delta$. The mean numbers of atoms arriving at the detector A or detector B are given by

$$_N<\Psi|N_A|\Psi>_N = j\cdot\Delta t\cdot\sin^2\frac{\varphi_{\alpha\beta}}{2}\cdot\frac{\Delta t}{T} \qquad (10)$$

$$_N<\Psi|N_B|\Psi>_N = j\cdot\Delta t\cdot\cos^2\frac{\varphi_{\alpha\beta}}{2}\cdot\frac{\Delta t}{T}. \qquad (11)$$

As for the multi-period measurement, we have to replace $\Delta t$ by $m\Delta t$, $T$ by $mT$, where $m$ is the number of periods. Seen from Equ.(10) and Equ.(11), the vibrant factor corrections are independent of the number of measurement periods and invariant. If there is not the vibrations of the atomic branches, the mean numbers of atoms arriving at the detectors A or detector B are written as[23]

$$_N<\Psi|N_A|\Psi>_{N0} = j\cdot\Delta t\cdot\sin^2\frac{\varphi_{\alpha\beta}}{2} \qquad (12)$$

$$_N\langle \Psi | N_B | \Psi \rangle_{N0} = j \cdot \Delta t \cdot \cos^2 \frac{\varphi_{\alpha\beta}}{2}. \qquad (13)$$

It is obvious that we can evaluate the frequency of the atomic branches after comparing the modified Equ.(10) and Equ.(11) with Equ.(12) and Equ.(13), the mean numbers of atoms arriving at the detectors have nothing to do with the absolute amplitude, so the vibrations, even micro-amplitude vibrations, are also verified by the measurement of the mean numbers of atoms arriving at detectors.

Without considering the harmonic vibrations of the two atomic branches, when the beams intensity is so low that there is only one atom at a time within a single coherence length, the quantum noise fluctuations of the detectors are given by[23]

$$\langle \Delta N_{A,B} \rangle_0 = \frac{\sqrt{N}}{2} \sin \varphi_{\alpha\beta} \qquad (14)$$

with $N = j \cdot \Delta t$. The relative quantum noise fluctuations are written as

$$\frac{\langle \Delta N_A \rangle_0}{\langle N_A \rangle_0} = \frac{\cot \frac{\varphi_{\alpha\beta}}{2}}{\sqrt{N}}$$

$$\frac{\langle \Delta N_B \rangle_0}{\langle N_B \rangle_0} = \frac{\tan \frac{\varphi_{\alpha\beta}}{2}}{\sqrt{N}}.$$

If we set $\varphi_{\alpha\beta} \equiv k(l_\alpha - l_\beta) \to 0$, we obtain $\frac{\langle \Delta N_B \rangle_0}{\langle N_B \rangle_0} \to 0$. For the detector B, the mean number correction of atoms due to the vibrations of the branches i.e. Equ.(11) is very obvious. That is to say, the effect of the harmonic vibrations of the two branches is not submerged in the quantum noise fluctuation of the detector B,.

**IV. Summary**

It is shown that detecting the vibrations of micro amplitudes and low frequencies is possible by measuring the mean numbers variations of atoms arriving at the detectors during a short time $\Delta t$ in an atomic Mach-Zehnder interferometer. In the condition of low frequencies the vibrant factor $F$ is proposed. The mean numbers of atoms arriving at the detectors are corrected by the vibrant factor $F$. The vibrant factor $F$ depends on the relative vibrant displacement and the initial phase, rather than the absolute vibrant amplitude. The vibrant factor $F$ increases with the increase of the initial phase. The frequency of the atomic branches can be evaluated from the measurement of the mean numbers variations of atoms. The quantum noise fluctuations of the detectors are discussed, the effect of the two branches vibrations on the mean numbers of atoms is not submerged in the quantum noise fluctuations of the detector B. The present results maybe provide a new detection principle for a gravitational wave detector. In the weak gravitational field approximation, spacetime is curved in a linearized theory. The simplest solution is a monochromatic plane wave solution, which amplitudes have two independent modes of polarization[25]. The deformed motions of a ring of test particles induced by the plane wave are quadrupole moment vibrations, up to now the effects of quadrupole moment vibrations of the beams on the mean numbers of atoms arriving at the detectors are still an open problem.